\documentclass[twocolumn,aps,prl,superscriptaddress,showpacs,preprintnumbers,amsmath,amssymb]{revtex4-1}
\usepackage[latin9]{inputenc}
\usepackage{amstext}
\usepackage{graphicx}

\makeatletter

\providecommand{\tabularnewline}{\\}

\usepackage{graphicx}
\usepackage{dcolumn}
\usepackage{accents}
\graphicspath{{ps}}

\renewcommand{\arraystretch}{1.1}
\renewcommand{\[}{\begin{equation}} 
\renewcommand{\]}{\end{equation}}

\newcommand\varpm{\mathbin{\vcenter{\hbox{%
  \oalign{\hfil$\scriptstyle+$\hfil\cr
          \noalign{\kern-.3ex}
          $\scriptscriptstyle({-})$\cr}%
}}}}
\newcommand\varmp{\mathbin{\vcenter{\hbox{%
  \oalign{$\scriptstyle({+})$\cr
          \noalign{\kern-.3ex}
          \hfil$\scriptscriptstyle{-}$\hfil\cr}%
}}}}

\def\pM{\mathrel{\raise 3pt \hbox{\tiny$\left(\right.$}\! 
                 \raise 2pt \hbox{+}
                 \settowidth {\dimen03} {+}
                 \hskip-\dimen03
                 \raise -2.4pt \hbox {$-$}
                 \!\raise 3pt \hbox{\tiny$\left.\right)$}}}
\def\mP{\mathrel{\raise 5pt \hbox{\tiny$\left(\right.$}\! 
                 \raise 4pt \hbox{$-$}
                 \settowidth {\dimen03} {$-$}
                 \hskip-\dimen03
                 \raise -2.4pt \hbox {$+$}
                 \!\raise 5pt \hbox{\tiny$\left.\right)$}}}

\makeatother

\begin{document}

\title{Search for $CP$ violation with kinematic asymmetries in the $D^{0}\rightarrow K^{+}K^{-}\pi^{+}\pi^{-}$
decay}


\noaffiliation
\affiliation{University of the Basque Country UPV/EHU, 48080 Bilbao}
\affiliation{Beihang University, Beijing 100191}
\affiliation{Brookhaven National Laboratory, Upton, New York 11973}
\affiliation{Budker Institute of Nuclear Physics SB RAS, Novosibirsk 630090}
\affiliation{Faculty of Mathematics and Physics, Charles University, 121 16 Prague}
\affiliation{University of Cincinnati, Cincinnati, Ohio 45221}
\affiliation{Deutsches Elektronen--Synchrotron, 22607 Hamburg}
\affiliation{University of Florida, Gainesville, Florida 32611}
\affiliation{Key Laboratory of Nuclear Physics and Ion-beam Application (MOE) and Institute of Modern Physics, Fudan University, Shanghai 200443}
\affiliation{Justus-Liebig-Universit\"at Gie\ss{}en, 35392 Gie\ss{}en}
\affiliation{Gifu University, Gifu 501-1193}
\affiliation{II. Physikalisches Institut, Georg-August-Universit\"at G\"ottingen, 37073 G\"ottingen}
\affiliation{SOKENDAI (The Graduate University for Advanced Studies), Hayama 240-0193}
\affiliation{Gyeongsang National University, Chinju 660-701}
\affiliation{Hanyang University, Seoul 133-791}
\affiliation{University of Hawaii, Honolulu, Hawaii 96822}
\affiliation{High Energy Accelerator Research Organization (KEK), Tsukuba 305-0801}
\affiliation{J-PARC Branch, KEK Theory Center, High Energy Accelerator Research Organization (KEK), Tsukuba 305-0801}
\affiliation{Forschungszentrum J\"{u}lich, 52425 J\"{u}lich}
\affiliation{IKERBASQUE, Basque Foundation for Science, 48013 Bilbao}
\affiliation{Indian Institute of Technology Bhubaneswar, Satya Nagar 751007}
\affiliation{Indian Institute of Technology Guwahati, Assam 781039}
\affiliation{Indian Institute of Technology Hyderabad, Telangana 502285}
\affiliation{Indian Institute of Technology Madras, Chennai 600036}
\affiliation{Indiana University, Bloomington, Indiana 47408}
\affiliation{Institute of High Energy Physics, Chinese Academy of Sciences, Beijing 100049}
\affiliation{Institute of High Energy Physics, Vienna 1050}
\affiliation{INFN - Sezione di Napoli, 80126 Napoli}
\affiliation{INFN - Sezione di Torino, 10125 Torino}
\affiliation{Advanced Science Research Center, Japan Atomic Energy Agency, Naka 319-1195}
\affiliation{J. Stefan Institute, 1000 Ljubljana}
\affiliation{Institut f\"ur Experimentelle Teilchenphysik, Karlsruher Institut f\"ur Technologie, 76131 Karlsruhe}
\affiliation{Kennesaw State University, Kennesaw, Georgia 30144}
\affiliation{King Abdulaziz City for Science and Technology, Riyadh 11442}
\affiliation{Department of Physics, Faculty of Science, King Abdulaziz University, Jeddah 21589}
\affiliation{Kitasato University, Sagamihara 252-0373}
\affiliation{Korea Institute of Science and Technology Information, Daejeon 305-806}
\affiliation{Korea University, Seoul 136-713}
\affiliation{Kyungpook National University, Daegu 702-701}
\affiliation{LAL, Univ. Paris-Sud, CNRS/IN2P3, Universit\'{e} Paris-Saclay, Orsay}
\affiliation{\'Ecole Polytechnique F\'ed\'erale de Lausanne (EPFL), Lausanne 1015}
\affiliation{P.N. Lebedev Physical Institute of the Russian Academy of Sciences, Moscow 119991}
\affiliation{Faculty of Mathematics and Physics, University of Ljubljana, 1000 Ljubljana}
\affiliation{Ludwig Maximilians University, 80539 Munich}
\affiliation{Luther College, Decorah, Iowa 52101}
\affiliation{Malaviya National Institute of Technology Jaipur, Jaipur 302017}
\affiliation{University of Malaya, 50603 Kuala Lumpur}
\affiliation{University of Maribor, 2000 Maribor}
\affiliation{Max-Planck-Institut f\"ur Physik, 80805 M\"unchen}
\affiliation{School of Physics, University of Melbourne, Victoria 3010}
\affiliation{University of Mississippi, University, Mississippi 38677}
\affiliation{University of Miyazaki, Miyazaki 889-2192}
\affiliation{Moscow Physical Engineering Institute, Moscow 115409}
\affiliation{Moscow Institute of Physics and Technology, Moscow Region 141700}
\affiliation{Graduate School of Science, Nagoya University, Nagoya 464-8602}
\affiliation{Kobayashi-Maskawa Institute, Nagoya University, Nagoya 464-8602}
\affiliation{Universit\`{a} di Napoli Federico II, 80055 Napoli}
\affiliation{Nara Women's University, Nara 630-8506}
\affiliation{National Central University, Chung-li 32054}
\affiliation{National United University, Miao Li 36003}
\affiliation{Department of Physics, National Taiwan University, Taipei 10617}
\affiliation{H. Niewodniczanski Institute of Nuclear Physics, Krakow 31-342}
\affiliation{Niigata University, Niigata 950-2181}
\affiliation{Novosibirsk State University, Novosibirsk 630090}
\affiliation{Osaka City University, Osaka 558-8585}
\affiliation{Pacific Northwest National Laboratory, Richland, Washington 99352}
\affiliation{Panjab University, Chandigarh 160014}
\affiliation{Peking University, Beijing 100871}
\affiliation{University of Pittsburgh, Pittsburgh, Pennsylvania 15260}
\affiliation{Theoretical Research Division, Nishina Center, RIKEN, Saitama 351-0198}
\affiliation{University of Science and Technology of China, Hefei 230026}
\affiliation{Seoul National University, Seoul 151-742}
\affiliation{Showa Pharmaceutical University, Tokyo 194-8543}
\affiliation{Soongsil University, Seoul 156-743}
\affiliation{University of South Carolina, Columbia, South Carolina 29208}
\affiliation{Stefan Meyer Institute for Subatomic Physics, Vienna 1090}
\affiliation{Sungkyunkwan University, Suwon 440-746}
\affiliation{School of Physics, University of Sydney, New South Wales 2006}
\affiliation{Department of Physics, Faculty of Science, University of Tabuk, Tabuk 71451}
\affiliation{Tata Institute of Fundamental Research, Mumbai 400005}
\affiliation{Department of Physics, Technische Universit\"at M\"unchen, 85748 Garching}
\affiliation{Toho University, Funabashi 274-8510}
\affiliation{Department of Physics, Tohoku University, Sendai 980-8578}
\affiliation{Earthquake Research Institute, University of Tokyo, Tokyo 113-0032}
\affiliation{Department of Physics, University of Tokyo, Tokyo 113-0033}
\affiliation{Tokyo Institute of Technology, Tokyo 152-8550}
\affiliation{Tokyo Metropolitan University, Tokyo 192-0397}
\affiliation{Virginia Polytechnic Institute and State University, Blacksburg, Virginia 24061}
\affiliation{Wayne State University, Detroit, Michigan 48202}
\affiliation{Yamagata University, Yamagata 990-8560}
\affiliation{Yonsei University, Seoul 120-749}
  \author{J.~B.~Kim}\affiliation{Korea University, Seoul 136-713} 
  \author{E.~Won}\email[Corresponding author:~~]{eunil@hep.korea.ac.kr}\affiliation{Korea University, Seoul 136-713} 
  \author{I.~Adachi}\affiliation{High Energy Accelerator Research Organization (KEK), Tsukuba 305-0801}\affiliation{SOKENDAI (The Graduate University for Advanced Studies), Hayama 240-0193} 
  \author{H.~Aihara}\affiliation{Department of Physics, University of Tokyo, Tokyo 113-0033} 
  \author{S.~Al~Said}\affiliation{Department of Physics, Faculty of Science, University of Tabuk, Tabuk 71451}\affiliation{Department of Physics, Faculty of Science, King Abdulaziz University, Jeddah 21589} 
  \author{D.~M.~Asner}\affiliation{Brookhaven National Laboratory, Upton, New York 11973} 
  \author{H.~Atmacan}\affiliation{University of South Carolina, Columbia, South Carolina 29208} 
  \author{V.~Aulchenko}\affiliation{Budker Institute of Nuclear Physics SB RAS, Novosibirsk 630090}\affiliation{Novosibirsk State University, Novosibirsk 630090} 
  \author{T.~Aushev}\affiliation{Moscow Institute of Physics and Technology, Moscow Region 141700} 
  \author{V.~Babu}\affiliation{Tata Institute of Fundamental Research, Mumbai 400005} 
% ...
\collaboration{The Belle Collaboration}

\begin{abstract}
We search for $CP$ violation in the singly-Cabibbo-suppressed decay
$D^{0}\rightarrow K^{+}K^{-}\pi^{+}\pi^{-}$ using data corresponding
to an integrated luminosity of $988\text{ }{\rm fb}^{-1}$ collected
by the Belle detector at the KEKB $e^{+}e^{-}$ collider. We measure
a set of five kinematically dependent $CP$ asymmetries, of which
four asymmetries are measured for the first time. The set of asymmetry
measurements can be sensitive to $CP$ violation via interference
between the different partial-wave contributions to the decay and
performed on other pseudoscalar decays. We find no evidence of $CP$
violation.
\end{abstract}


\maketitle
\tighten

{\renewcommand{\thefootnote}{\fnsymbol{footnote}}}
\setcounter{footnote}{0}

Charge-conjugation and parity $(CP)$ symmetry violation has been
observed in various weak decays involving strange and beauty quarks
\cite{PDG} and is well described in the standard model (SM) by the
Cabibbo-Kobayashi-Maskawa matrix \cite{CKMMatrix}. But the magnitude
of $CP$ violation in the SM is too small to explain the baryon asymmetry
in the visible universe \cite{Baryogenesis}. Therefore, the search
for additional processes that violate $CP$ symmetry, which are not
described by the SM, is of great interest to explain the matter-dominant
universe. $CP$ violation in the charm sector is expected to be small,
less than $\mathcal{O}\left(10^{-3}\right)$ in the SM \cite{CPCharmOrder1,CPCharmOrder2},
which makes it an excellent probe for $CP$ violation beyond that
of the SM \cite{PDG}. 

$CP$ violation in the singly-Cabibbo-suppressed decay $D^{0}\rightarrow K^{+}K^{-}\pi^{+}\pi^{-}$
was searched for using $\hat{T}$-odd correlations \cite{FocusSearch,BabarSearch,LHCbSearch},
where $\hat{T}$ reverses the direction of momenta and spin, which
is different from the usual time reversal operator $T$ \cite{ProbeCPSystematic}.
No $CP$ violation is observed up to now, but the $\hat{T}$-odd correlation
measured may be weakly sensitive to $CP$ violation in this decay
\cite{ProbeCPSystematic}. In this paper, we report the first measurement
of a set of $CP$-violating kinematic asymmetries in $D^{0}\rightarrow K^{+}K^{-}\pi^{+}\pi^{-}$
decays. The set of kinematic asymmetries probes the rich variety of
interfering contributions in a decay, which can be sensitive to non-SM
$CP$-violating phases. 

Assuming $CPT$ symmetry, we construct a $CP$-violating asymmetry
by comparing amplitudes of the decay with their $CP$-conjugate amplitudes.
Amplitudes of the decay can be extracted from $\mathcal{A}_{X}$,
which we define as 
\begin{equation}
\mathcal{A}_{X}\equiv\frac{\Gamma\left(X>0\right)-\Gamma\left(X<0\right)}{\Gamma\left(X>0\right)+\Gamma\left(X<0\right)},
\end{equation}
where $X$ is a kinematic variable, such as the vector triple product
of the final-state momenta used in Ref. \cite{FocusSearch,BabarSearch,LHCbSearch},
$\Gamma\left(X>0\right)$ is the rate for $D^{0}$ decays in which
$X>0$; and $\Gamma\left(X<0\right)$, for $D^{0}$ decays in which
$X<0$. The $CP$-conjugated amplitudes can be extracted similarly
for $\bar{D}^{0}$ decays using $\bar{X}$. We can then define our
$CP$-violating kinematic asymmetry as 
\begin{equation}
a_{X}^{CP}\equiv\frac{1}{2}\left(\mathcal{A}_{X}-\eta_{X}^{CP}\bar{\mathcal{A}}_{\bar{X}}\right),
\end{equation}
where $\eta_{X}^{CP}$ is a $CP$ eigenvalue specific to $X$.

\begin{figure}
\begin{centering}
\includegraphics[width=1\columnwidth]{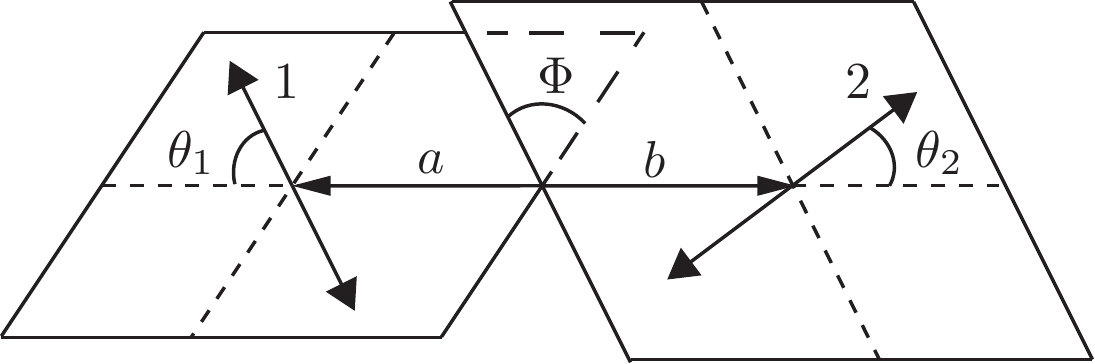}
\par\end{centering}
\caption{Diagram showing the helicity angles $\theta_{1}$, $\theta_{2}$ and
$\Phi$. \label{fig:CabibboMaksymowiczAngles}}
\end{figure}

We measure a set of kinematic asymmetries for five different $X$,
where four asymmetries are measured for the first time and one asymmetry
is proportional to the $\hat{T}$-odd correlation using the vector
triple product of the final-state momenta, which has been measured
previously \cite{FocusSearch,BabarSearch,LHCbSearch}. The set can
be sensitive to $CP$ violation in the interference between the $S$-wave
and $P$-wave production of the $K^{+}K^{-}$ and $\pi^{+}\pi^{-}$
pairs in the $D^{0}\rightarrow K^{+}K^{-}\pi^{+}\pi^{-}$ decay, where
the process of a quasi-two-body decay to a dikaon system and dipion
system contributes to over 40\% of the decay rate \cite{PDG}. It
covers the asymmetries that can be measured without considering the
mass of the intermediate particles. The kinematic variables are constructed
from the angles $\theta_{1}$, $\theta_{2}$, and $\Phi$, which are
shown in Fig. \ref{fig:CabibboMaksymowiczAngles}. The $\theta_{1}$
is the angle between the $K^{+}$ momentum and the direction opposite
to that of the $D^{0}$ momentum in the center-of-mass (CM) frame
of the $K^{+}K^{-}$ system. The $\theta_{2}$ is defined in the same
way as $\theta_{1}$ substituting $K^{+}$ with $\pi^{+}$ and $K^{+}K^{-}$
with $\pi^{+}\pi^{-}$. The $\Phi$ is the angle between the decay
planes of the $K^{+}K^{-}$ and $\pi^{+}\pi^{-}$ pairs in the CM
frame of $D^{0}$. Three kinematic variables have $\eta_{X}^{CP}=-1$:
$\sin2\Phi$, $\cos\theta_{1}\cos\theta_{2}\sin\Phi$, and $\sin\Phi$;
the last variable is proportional to the vector triple product of
the final-state momenta. The remaining two kinematic variables have
$\eta_{X}^{CP}=+1$: $\cos\Phi$ and $\cos\theta_{1}\cos\theta_{2}\cos\Phi$.
The kinematic asymmetries where $\eta_{X}^{CP}$ is $-1$, commonly
known as $\hat{T}$-odd correlations, are dependent on the imaginary
part of the interference of amplitudes for production of the $K^{+}K^{-}$
and $\pi^{+}\pi^{-}$ states in different spin configurations \cite{DefineDecayRateRosner,DefineDecayRateLondon,DefineDecayRateWithMass,DefineDecayRateWithMass2}.
The asymmetries where $\eta_{X}^{CP}$ is $+1$, are dependent on
the real part of the interference of amplitudes. Both types of asymmetries
are nonzero in the case of $CP$ violation. This set is measured for
the first time for any four-body final state; these measurements can
be performed for any other pseudoscalar meson that decays to four
pseudoscalar mesons.

This analysis uses the data sample recorded by the Belle detector
\cite{Belle2} at the $e^{+}e^{-}$ asymmetric-energy collider KEKB
\cite{KEKB2}, where the CM energy of the collisions was varied from
the mass of the $\Upsilon(1\text{S})$ resonance up to that of the
$\Upsilon(6\text{S})$ resonance. The total data sample corresponds
to an integrated luminosity of 988 $\text{fb}^{-1}$ \cite{Luminosity}.

Monte Carlo (MC) samples are used to devise the selection criteria,
identify the different sources of background, model the data, validate
the fit procedure, and determine systematic uncertainties. Inclusive
MC samples were generated with \texttt{EvtGen} \cite{EVTGEN}, where
the number of generated events corresponds to six times the integrated
luminosity of the data sample. The detector response was simulated
with \texttt{GEANT3} \cite{Geant}. To simulate the effect of beam-induced
background, the generated events have data solely due to the beam
backgrounds overlaid.

Since the final state is self-conjugate, the flavor of the $D^{0}$
mesons is identified by reconstructing the decay chains $D^{*+}\rightarrow D^{0}\pi_{\text{s}}^{+}$,
with $D^{0}$ decaying into $K^{+}K^{-}\pi^{+}\pi^{-}$, where $\pi_{\text{s}}^{+}$
is referred to as the slow pion. Here, and elsewhere in this paper,
charge-conjugate states are implied unless stated explicitly otherwise.

Using MC simulated data, we developed the selection criteria to maximize
a figure of merit of $S/\sqrt{S+B}$, where $S$ is the signal yield
and $B$ is the background yield in a signal enhanced region, which
is defined to be within $1.5$ $\text{MeV}/c^{2}$ of the known $D^{0}$
mass \cite{PDG} and within $0.25$ $\text{MeV}/c^{2}$ of the known
mass difference $\left(\Delta m\right)$ between the $D^{*+}$ candidate
and its daughter $D^{0}$ \cite{PDG}.

We select charged tracks that originate from close to the $e^{+}e^{-}$
interaction point (IP) by requiring the impact parameters to be less
than 4 cm in the beam direction and 2 cm in the plane transverse to
the beam direction. To ensure the tracks are well reconstructed, we
require they each have a transverse momentum greater than 0.1 $\text{GeV}/c$
and at least two associated hits in the silicon vertex detector in
both the beam direction and azimuthal direction. Charged tracks are
identified as pions or kaons depending on the ratio of particle identification
likelihoods $\mathcal{L}_{K}/\left(\mathcal{L}_{K}+\mathcal{L}_{\pi}\right)$,
which are constructed from information recorded by the central drift
chamber, time-of-flight scintillation counters, and aerogel threshold
Cherenkov counter. We identify a charged track as a kaon when this
ratio is above 0.6; otherwise it is assumed to be a pion. The kaon
and pion identification efficiencies are typically over 80\%, and
the misidentification probabilities are below 10\% \cite{PID}.

We form a $D^{0}$ candidate from each combination of two oppositely
charged kaon tracks and two oppositely charged pion tracks. We require
each $D^{0}$ candidate have an invariant mass within $30\text{ MeV/}c^{2}$
of the known $D^{0}$ mass \cite{D0K0mass,PDG}, where the range is
larger than 7 times the mass resolution of the reconstructed $D^{0}$
candidate, and a momentum in the CM frame greater than 1.8 $\text{GeV}/c$.
For each surviving candidate, we perform a vertex- and mass-constrained
fit to the kaons and pions; we require the vertex fit to have a probability
greater than 0.1\%. We also perform a fit where each $D^{0}$ candidate
is fit under the hypothesis that the trajectory of the candidate originates
from the IP and require the fit to have a probability greater than
0.005\%. 

To veto the Cabibbo-favored $D^{0}\rightarrow K^{+}K^{-}K_{\text{S}}^{0}$
decays, we remove $D^{0}$ candidates whose daughter pion pairs have
invariant masses within $12.05\text{ }\text{MeV}/c^{2}$ of the known
$K_{\text{S}}^{0}$ mass \cite{PDG}, which is five times the mass
resolution of the reconstructed $K_{\text{S}}^{0}$ candidate. 

We form each combination of a positively charged pion track and $D^{0}$
candidate into a $D^{*+}$ candidate and perform a vertex fit on the
pion, where the fit is constrained to the intersection of the $D^{0}$
candidate trajectory with the IP region. We require each $D^{*+}$
candidate have a momentum in the CM frame greater than $2.5\text{ }\text{GeV}/c$.
We also require $\Delta m$ to be within $_{-5.9}^{+7.6}\text{ MeV}/c^{2}$
of the known $\Delta m$ \cite{PDG}, where the lower limit corresponds
to the known $\pi^{\pm}$ mass.

In the signal region, 8.1\% of events have multiple $D^{*+}$ and/or
$D^{*-}$ candidates, while the average multiple candidates per event
is 1.1, which is comparable with Ref.~\cite{MultipleCandBelle}.
We select either a $D^{*+}$ or $D^{*-}$ candidate for each event,
based on the smallest $\chi^{2}$ for the $D^{0}$ mass fit. If there
are multiple $D^{*+}$ and/or $D^{*-}$ candidates formed with this
$D^{0}$, we select the one whose $\pi_{\text{s}}^{+}$ or $\pi_{\text{s}}^{-}$
has the smallest impact parameter in the transverse plane. Studies
with the MC sample indicate that 93\% of the multiple-candidate events
are correctly selected. The efficiency for the $D^{0}\rightarrow K^{+}K^{-}\pi^{+}\pi^{-}$
decay with the stated selections is 11\%. A total of 474,971 events
are reconstructed from the data sample.

After all selection criteria, our data sample contains events that
fall into four different categories: correctly reconstructed $D^{0}$
mesons coming from correctly reconstructed $D^{*+}$ mesons, which
we call signal events; events with correctly reconstructed $D^{0}$
mesons coming from misreconstructed $D^{*+}$ candidates, which we
call random-$\pi_{\text{s}}$ events; events with a partially reconstructed
$D^{0}$ candidate and the $\pi_{s}^{+}$ from a $D^{*+}$, which
we call partial-$D^{*}$ events, which has a small peak in the signal
region of $\Delta m$; and events with both $D^{0}$ and $D^{*+}$
candidates misreconstructed, which we call combinatorial events. Our
selection criteria rejects over 99\% of events with $D_{s}^{+}\rightarrow K^{+}K^{-}\pi^{+}\pi^{+}\pi^{-}$,
which could be confused for our signal, leaving a negligible number
of such events.

We calculate the $CP$-violating kinematic asymmetry with the yield
of the signal events for each flavor of $D^{0}$ and each sign of
the relevant kinematic variable. To do this, we perform four separate
fits to the data for each kinematic variable. Each fit is a binned
two-dimensional extended maximum-likelihood fit to the reconstructed
$D^{0}$ mass and $\Delta m$. The data are binned into 200 equal-width
bins in each dimension. These additional requirements on $m(K^{+}K^{-}\pi^{+}\pi^{-})$
and $\Delta m$ have a negligible effect on the selection efficiency. 

One model is used for all fits. It contains components describing
signal, random $\pi_{\text{s}}$, partial-$D^{*}$, and combinatorial
events. The yield of each component is free in each fit, but parameters
governing the shapes of the components are fixed from a single fit
to all the data regardless of $D^{0}$ flavor and $X$.

The signal component is the product of a sum of bifurcated Gaussian
and Gaussian probability density functions (PDFs) for $m\left(K^{+}K^{-}\pi^{+}\pi^{-}\right)$
and a sum of Gaussian and JohnsonSU \cite{JohnsonSU} PDFs for $\Delta m$.
The combinatorial component is the product of a Chebyshev function
for $m\left(K^{+}K^{-}\pi^{+}\pi^{-}\right)$ and a threshold function
for $\Delta m$. The random-$\pi_{\text{s}}$ component is the product
of the signal shape for $m\left(K^{+}K^{-}\pi^{+}\pi^{-}\right)$
and the combinatorial shape for $\Delta m$. And the partial-$D^{*}$
component is the product of a Chebyshev function for $m\left(K^{+}K^{-}\pi^{+}\pi^{-}\right)$
and a Bifurcated Gaussian PDF for $\Delta m$, where the shape parameters
for the partial-$D^{*}$ component are fixed to those obtained from
a fit to an inclusive MC sample. The shape of the MC sample is validated
by comparing it to data in the sidebands of the $m\left(K^{+}K^{-}\pi^{+}\pi^{-}\right)$
distribution; the shapes are compatible. 

\begin{figure}
\begin{centering}
\includegraphics[width=0.5\columnwidth]{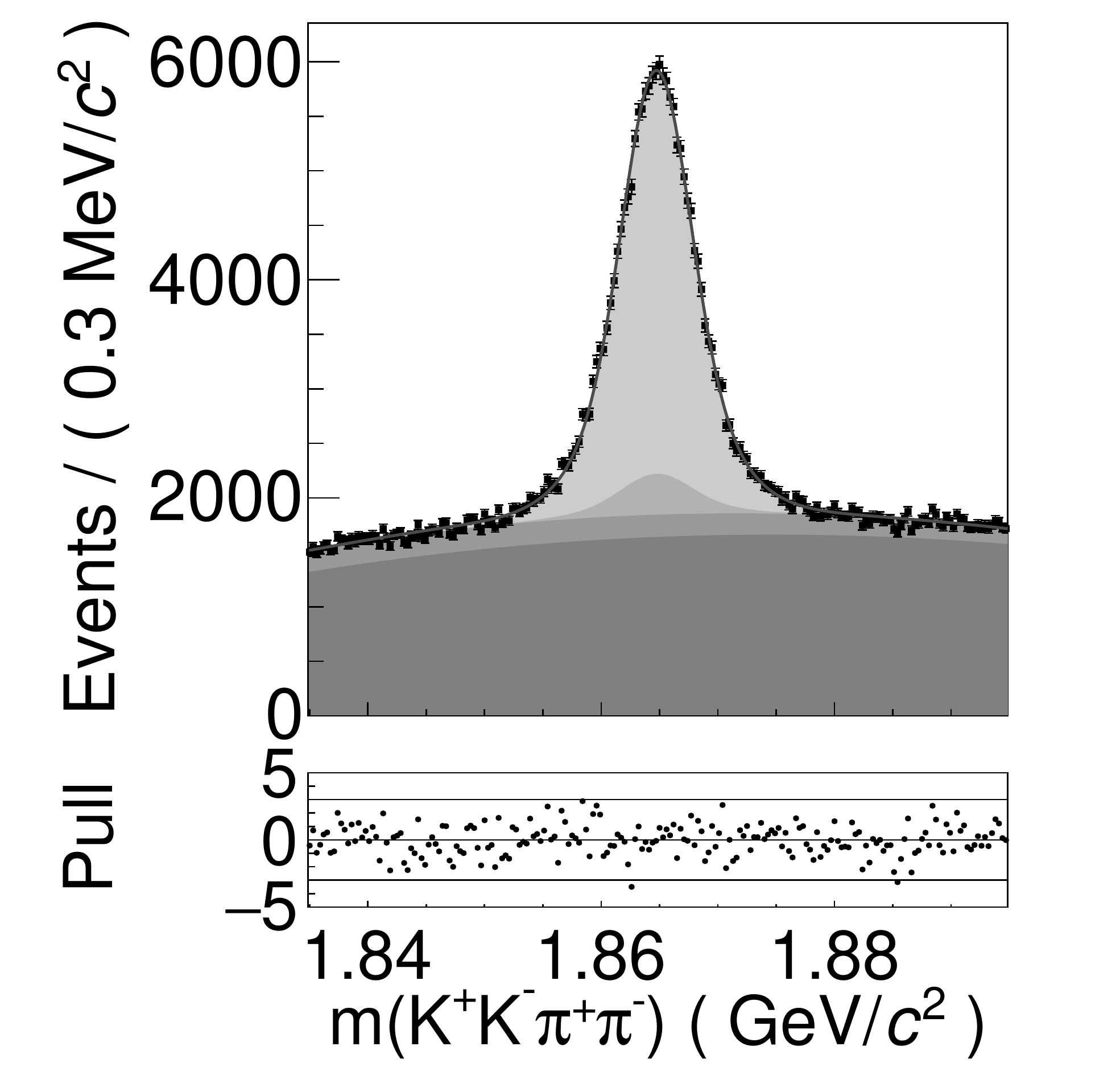}\includegraphics[width=0.5\columnwidth]{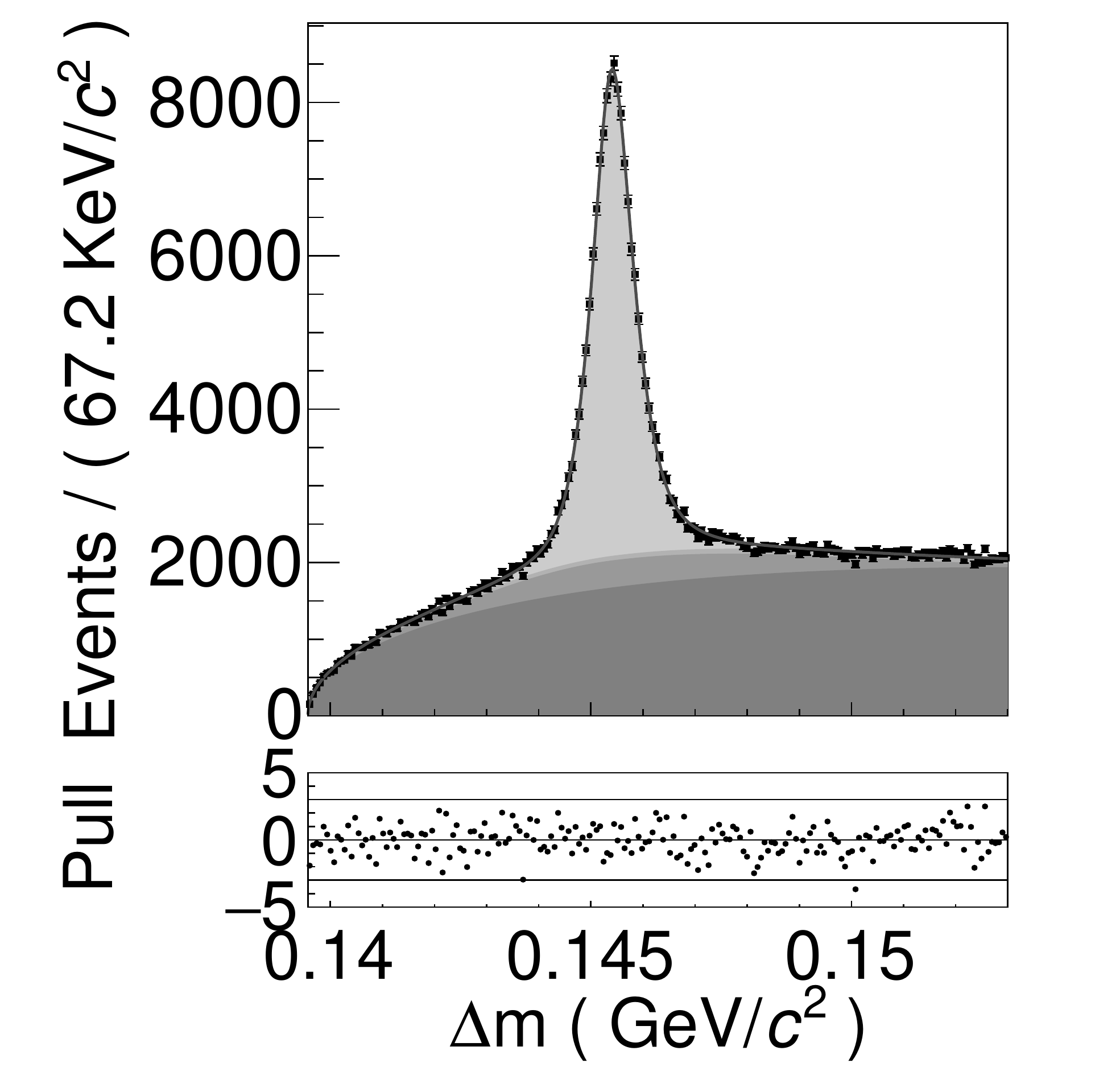}
\par\end{centering}
\caption{Data (black points) and fit results (shaded regions) for the fit to
all the data in projections of $m\left(K^{+}K^{-}\pi^{+}\pi^{-}\right)$
(left) and $\Delta m$ (right). The shaded regions are stacked upon
each other and show, from lowest to highest, the combinatorial, partial-$D^{*}$,
random-$\pi_{s}$, and signal components. The lower plots show the
pulls for the fit; the unlabeled horizontal lines indicate $\pm3$.
\label{fig:DistributionAndFit}}
\end{figure}

\begin{figure}
\begin{centering}
\includegraphics[width=0.5\columnwidth]{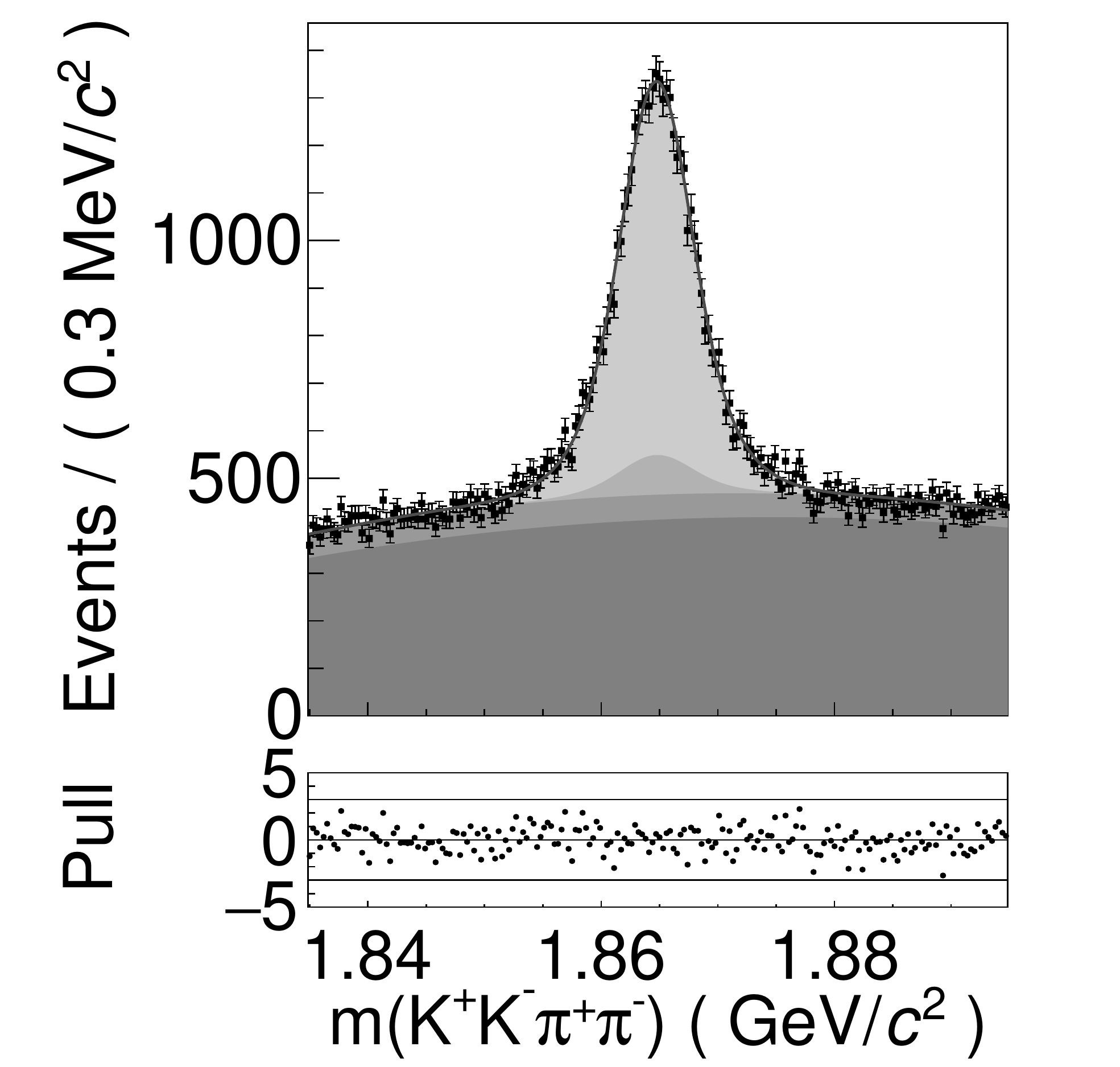}\includegraphics[width=0.5\columnwidth]{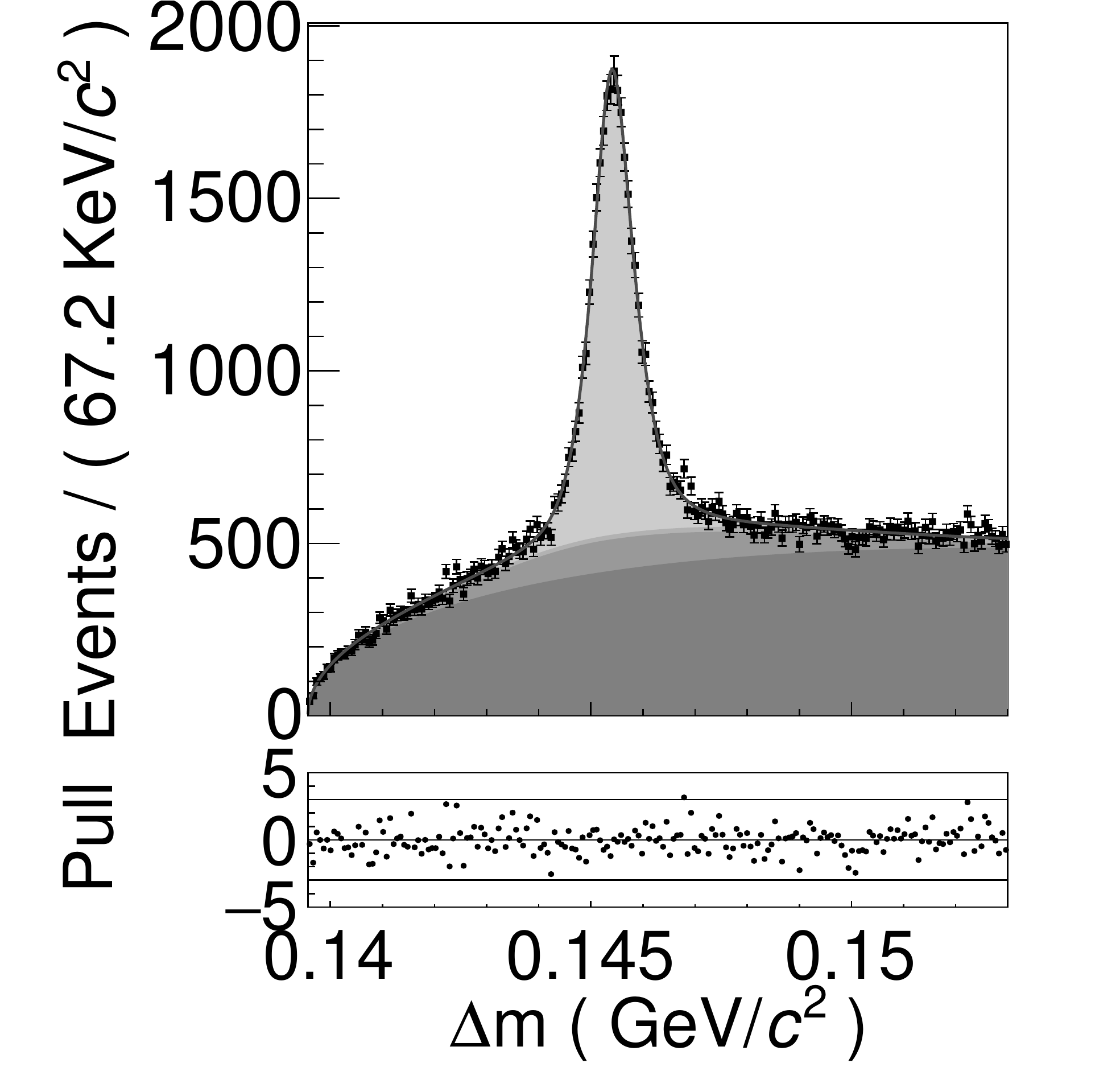}
\par\end{centering}
\begin{centering}
\includegraphics[width=0.5\columnwidth]{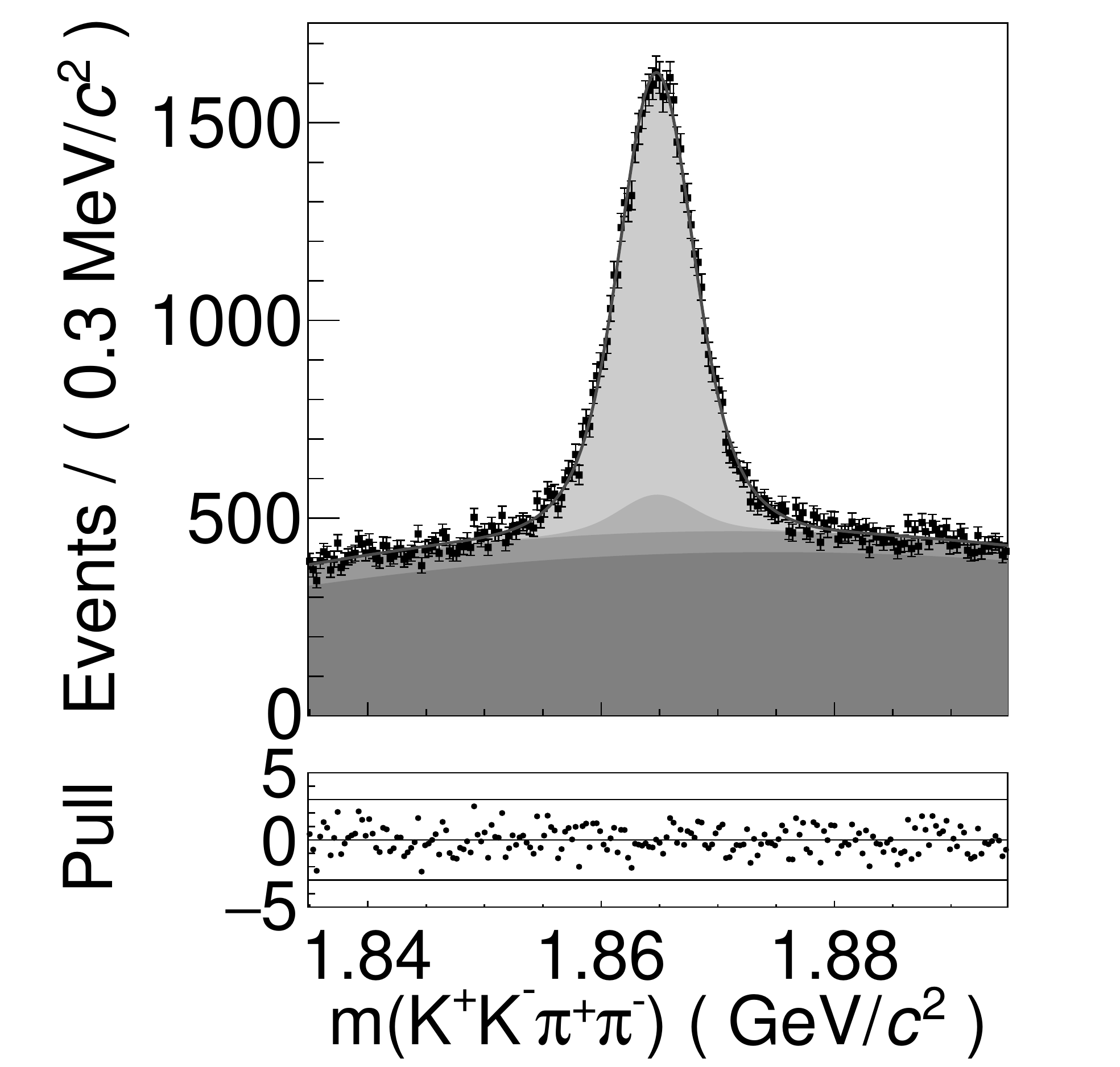}\includegraphics[width=0.5\columnwidth]{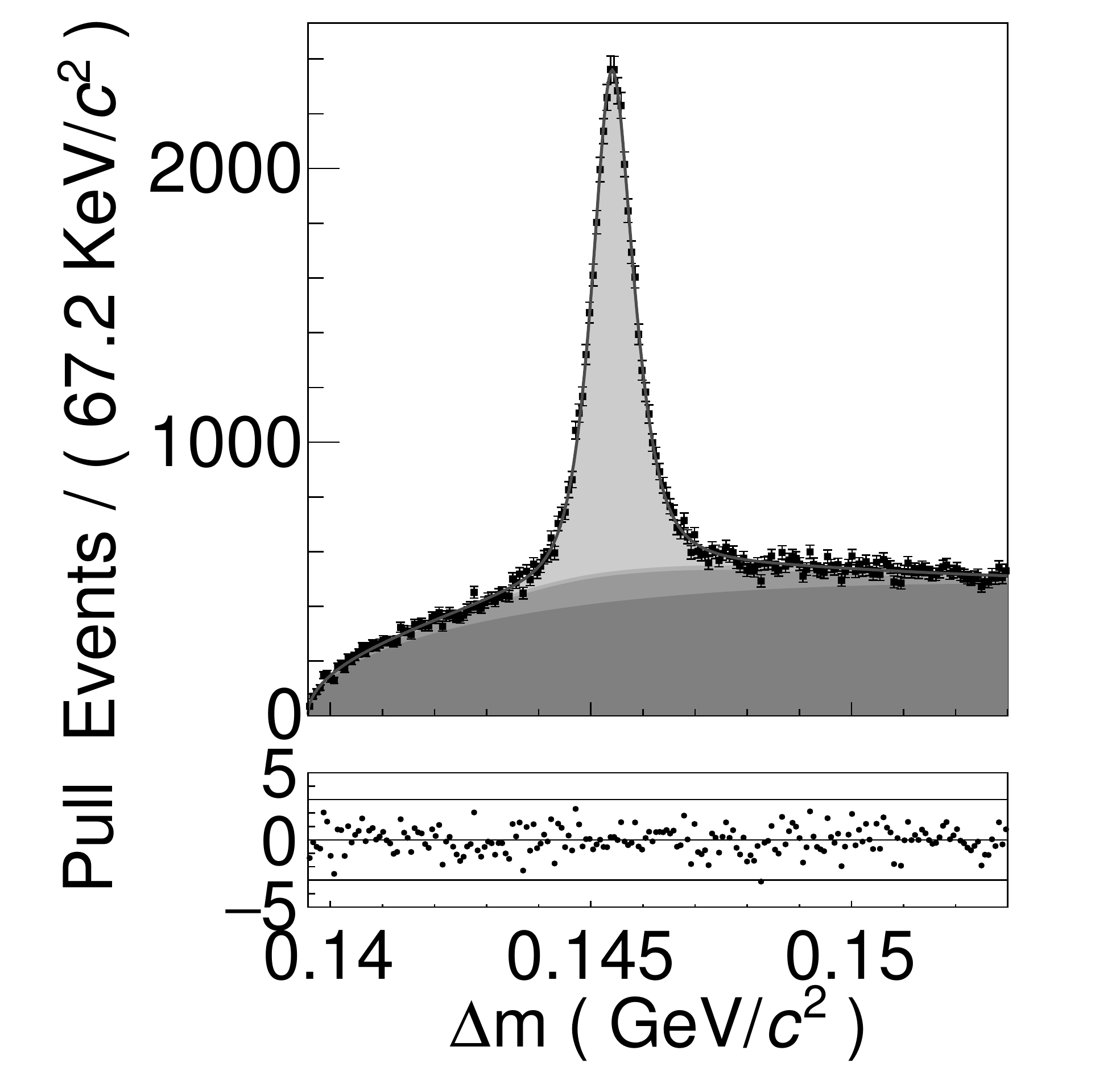}
\par\end{centering}
\caption{Two-dimensional fit results, distributions and pull of the data subsamples
projected on the observables $m\text{\ensuremath{\left(K^{+}K^{-}\pi^{+}\pi^{-}\right)}}$
(left) and $\Delta m$ (right). The distribution sequence follows
Fig. \ref{fig:DistributionAndFit}. Top histograms show the $D^{0}\left(\sin2\Phi>0\right)$
subsample while the bottom histograms show the $D^{0}\left(\sin2\Phi<0\right)$
subsample. \label{fig:2DSimFitDsp}}
\end{figure}

\renewcommand{\arraystretch}{1.3}

\begin{table*}
\caption{Fit results for the yield of signal in the subsamples of each kinematic
variable $X$. The uncertainties are statistical only.\label{tab:FitResults}}
\begin{ruledtabular}
\begin{centering}
\begin{tabular}{ccccc}
$X$ & $D^{0}\left(X>0\right)$  & $D^{0}\left(X<0\right)$ & $\bar{D}^{0}\left(\bar{X}>0\right)$  & $\bar{D}^{0}\left(\bar{X}<0\right)$ \tabularnewline
\hline 
$\cos\Phi$ & 21,913 $\pm$ 181 & 32,544 $\pm$ 216 & 21,657 $\pm$ 180 & 32,623 $\pm$ 216\tabularnewline
$\sin\Phi$ & 29,177 $\pm$ 205 & 25,277 $\pm$ 194 & 25,474 $\pm$ 194 & 28,800 $\pm$ 204\tabularnewline
$\sin2\Phi$ & 23,096 $\pm$ 187 & 31,355 $\pm$ 211 & 31,455 $\pm$ 211 & 22,805 $\pm$ 186\tabularnewline
$\cos\theta_{1}\cos\theta_{2}\cos\Phi$ & 31,065 $\pm$ 211 & 23,398 $\pm$ 188 & 30,963 $\pm$ 210 & 23,304 $\pm$ 187\tabularnewline
$\cos\theta_{1}\cos\theta_{2}\sin\Phi$ & 26,016 $\pm$ 196 & 28,441 $\pm$ 203 & 28,353 $\pm$ 203 & 25,919 $\pm$ 195\tabularnewline
\end{tabular}
\par\end{centering}
\end{ruledtabular}
\end{table*}
\renewcommand{\arraystretch}{1}

Figure~\ref{fig:DistributionAndFit} shows the results of the fit
to all the data, from which the shapes of all components are fixed
for all remaining fits; the model agrees well with the data, as can
be seen from the pulls, which are defined as the difference between
the data points and the model expectation divided by the expected
uncertainty. As an example of a set of fits used to determine the
$CP$-violating kinematic asymmetry, we show separate fit results
for positive and negative $\sin2\Phi$ for $D^{0}$ samples in Fig.~\ref{fig:2DSimFitDsp}.
The signal yields determined by the fits are given in Table~\ref{tab:FitResults}
for each $D^{0}$ flavor and kinematic variable sign.

\renewcommand{\arraystretch}{1.5}

\begin{table*}
\caption{Contributions to the systematic uncertainty (in per mille) for each
$CP$-violating kinematic asymmetry $a_{X}^{CP}$. \label{tab:systematics}}

\begin{ruledtabular}
\begin{centering}
\begin{tabular}{cccccc}
Effect & $a_{\cos\Phi}^{CP}$ & $a_{\sin\Phi}^{CP}$  & $a_{\sin2\Phi}^{CP}$  & $a_{\cos\theta_{1}\cos\theta_{2}\cos\Phi}^{CP}$  & $a_{\cos\theta_{1}\cos\theta_{2}\sin\Phi}^{CP}$ \tabularnewline
\hline 
Signal model PDF & 0.1 & 0.3 & 0.1 & 0.2 & 0.0\tabularnewline
Partial-$D^{*}$ model PDF & 0.1 & 0.1 & 0.2 & 0.2 & 0.0\tabularnewline
Combinatorial model PDF & 0.1 & 0.1 & 0.3 & 0.0 & 0.3\tabularnewline
Detector bias & 0.6 & 0.6 & 0.6 & 0.6 & 0.6\tabularnewline
Likelihood fit bias & 0.1 & 0.1 & 0.1 & 0.1 & 0.1\tabularnewline
\hline 
Total & 0.6 & 0.7 & 0.7 & 0.7 & 0.7\tabularnewline
\end{tabular}
\par\end{centering}
\end{ruledtabular}
\end{table*}

\renewcommand{\arraystretch}{1}

\renewcommand{\arraystretch}{1.3}

\begin{table}
\caption{$\mathcal{A}_{X}$ and $\mathcal{\bar{A}}_{X}$ (in per mille) for
each kinematic variable $X$. The uncertainties are statistical only.\label{tab:AmplitudeResults}}
\begin{ruledtabular}
\begin{centering}
\begin{tabular}{ccc}
$X$ & $\mathcal{A}_{X}$ & $\mathcal{\bar{A}}_{X}$\tabularnewline
\hline 
$\cos\Phi$ & -195.2 $\pm$ 5.1 & 202.0 $\pm$ 5.1\tabularnewline
$\sin\Phi$ & 71.6 $\pm$ 5.2 & 61.3 $\pm$ 5.2\tabularnewline
$\sin2\Phi$ & -151.7 $\pm$ 5.2 & -159.4 $\pm$ 5.1\tabularnewline
$\cos\theta_{1}\cos\theta_{2}\cos\Phi$ & 140.8 $\pm$ 5.1 & -141.2 $\pm$ 5.2\tabularnewline
$\cos\theta_{1}\cos\theta_{2}\sin\Phi$ & -44.5 $\pm$ 5.2 & -44.9 $\pm$ 5.2\tabularnewline
\end{tabular}
\par\end{centering}
\end{ruledtabular}
\end{table}
\renewcommand{\arraystretch}{1}

We perform several cross checks to validate our analysis: To study
the effect of the $D^{0}\rightarrow K^{*}\bar{K}^{*}$, where $K^{*}$
decays to $K^{+}\pi^{-}$, we recalculate the asymmetries including
a veto on $K^{*}$ and $\bar{K}^{*}$ that rejects the $D^{0}$ candidates
with a $K^{+}\pi^{-}$ pair and $K^{-}\pi^{+}$ pair of an invariant
mass within 80 $\text{MeV}/c^{2}$ of the known $K^{*}$ mass \cite{PDG},
which is twenty times the mass resolution of the reconstructed $K^{*}$
candidate. The recalculated asymmetries are consistent with the values
without the veto.

To study the effects from the best candidate selection, we recalculate
the asymmetries with no best candidate selection. The recalculated
asymmetries are consistent with those calculated including the best
candidate selection.

The detector resolution of the kinematic variables could affect the
asymmetries. We measure the fraction of cross-feed between signal
events with $X>0$ and $X<0$ using an MC sample that has a similar
shape to the data. The fraction of cross-feed is at the 1\% level,
making its effect negligible.

We estimate the effect of incorrectly assigning the flavor of the
$D^{0}$ using an MC sample that has a similar integrated luminosity
to the data. In the MC sample, incorrectly assigned events comprise
less than 0.01\% of the total number of events. Missassignment has
a negligible effect.

There could be an effect due to an efficiency difference depending
on the kinematic variable regions. Efficiencies depending on kinematic
variable regions are measured using a MC sample. We find that the
efficiency does not depend on the kinematic variables used to define
the asymmetries.

Several sources of systematic uncertainty are considered. Individual
uncertainties and the total systematic uncertainty are listed in Table~\ref{tab:systematics}.
The bias from the model PDF is estimated by changing the signal model,
partial-$D^{*}$ model, and combinatorial model. We change the signal
model and partial-$D^{*}$ model to products of one-dimensional Gaussian-kernel-estimated
PDFs \cite{PDFKEY} and the combinatorial model to a product of one-dimensional
PDFs obtained from an inclusive MC sample. The difference between
the measured values is assigned as a systematic uncertainty. 

The detector bias is estimated from a control sample of $D^{0}\rightarrow K^{-}\pi_{\text{low}}^{+}\pi^{-}\pi_{\text{high}}^{+}$
events, where momentum is used to differentiate between the $\pi_{{\rm high}}^{+}$
and $\pi_{{\rm low}}^{+}$. This decay is Cabibbo-favored in which
all kinematic asymmetries are expected to be much smaller than the
measurement precision \cite{CPCharmOrder1}. The kinematic variables
are calculated in the same way as for the $K^{+}K^{-}\pi^{+}\pi^{-}$
final state, substituting $K^{+}$ with $\pi_{\text{low}}^{+}$. The
kinematic asymmetries are found to be consistent with zero, and we
assign their statistical uncertainties as the systematic uncertainties
related to any detector bias. 

To assess whether there is a bias introduced by the likelihood fit
and to check the extraction of kinematic asymmetries from the two-dimensional
binned fit, we generate MC samples with different asymmetries and
compare the fit results with the generated values. The average difference
between the measured and generated value is assigned as a systematic
uncertainty.

The various sources of systematic uncertainty are independent of each
other. Therefore we estimate the total systematic uncertainty by summing
the uncertainties in quadrature. As a note, the kinematic asymmetries
are constructed such that they are insensitive to the intrinsic production
asymmetry \cite{LHCbSearch}.

The measured $\mathcal{A}_{X}$ and $\mathcal{\bar{A}}_{X}$ are listed
in Table~\ref{tab:AmplitudeResults} with statistical errors. As
in other experiments \cite{BabarSearch,LHCbSearch}, final state interaction
effects are observed with a similar amplitude for $\mathcal{A}_{\sin\Phi}$
and $\bar{\mathcal{A}}_{\sin\Phi}$. We find the $CP$-violating kinematic
asymmetries to be 
\begin{equation}
a_{\cos\Phi}^{CP}=\left(3.4\pm3.6\pm0.6\right)\times10^{-3},
\end{equation}
\begin{equation}
a_{\sin\Phi}^{CP}=\left(5.2\pm3.7\pm0.7\right)\times10^{-3},
\end{equation}
\begin{equation}
a_{\sin2\Phi}^{CP}=\left(3.9\pm3.6\pm0.7\right)\times10^{-3},
\end{equation}
\begin{equation}
a_{\cos\theta_{1}\cos\theta_{2}\cos\Phi}^{CP}=\left(-0.2\pm3.6\pm0.7\right)\times10^{-3},
\end{equation}
\begin{equation}
a_{\cos\theta_{1}\cos\theta_{2}\sin\Phi}^{CP}=\left(0.2\pm3.7\pm0.7\right)\times10^{-3},
\end{equation}
where the first and second uncertainties are statistical and systematic,
respectively. These results indicate that there is no $CP$ violation
within the statistical and systematic uncertainties for the interferences
between the S-wave and P-wave production of the $K^{+}K^{-}$ and
$\pi^{+}\pi^{-}$ pairs in this decay. No effects from new physics
models can be observed within the experimental uncertainties. With
more data from future experiments, it may be possible to measure the
$CP$ violation due to the SM in this decay.

In conclusion, we search for $CP$ violation in $D^{0}\rightarrow K^{+}K^{-}\pi^{+}\pi^{-}$
by measuring a set of five kinematic asymmetries. The set of measurements
can be sensitive to $CP$ violation via the rich variety of interference
between the different partial-wave contributions to the decay. It
can be performed on any other pseudoscalar meson that decays into
four pseudoscalar mesons. Four asymmetries are measured for the first
time. The set of $CP$-violating kinematic asymmetries is consistent
with $CP$ conservation and provide new constraints on new physics
models \cite{ProbeCPSystematic,CPCharmOrder1,DefineDecayRateLondon}.

We thank the KEKB group for excellent operation of the accelerator; the KEK cryogenics group for efficient solenoid operations; and the KEK computer group, the NII, and  PNNL/EMSL for valuable computing and SINET5 network support.   We acknowledge support from MEXT, JSPS and Nagoya's TLPRC (Japan); ARC (Australia); FWF (Austria); NSFC and CCEPP (China);  MSMT (Czechia); CZF, DFG, EXC153, and VS (Germany); DST (India); INFN (Italy);  MOE, MSIP, NRF, RSRI, FLRFAS project and GSDC of KISTI and KREONET/GLORIAD (Korea); MNiSW and NCN (Poland); MSHE, grant 14.W03.31.0026 (Russia); ARRS (Slovenia); IKERBASQUE and MINECO (Spain);  SNSF (Switzerland); MOE and MOST (Taiwan); and DOE and NSF (USA). E.~Won is partially supported by NRF-2017R1A2B3001968.


\bibliographystyle{elsTest}
\bibliography{d0_kkpipi_pub}

\end{document}